\documentclass[pre, reprint, superscriptaddress, amsmath, amssymb, aps]{revtex4-2}

\usepackage{graphicx}
\usepackage{dcolumn}
\usepackage{bm}
\usepackage{booktabs}

\usepackage{lmodern}
\usepackage{keyval}
\makeatletter
\providecommand{\pickup@font}{}
\makeatother

\usepackage[english]{babel}
\newcommand{\hyph}{\babelhyphen{hard}}

\usepackage[protrusion=true,expansion=true]{microtype}
\microtypesetup{activate=true}

\usepackage[normalem]{ulem}
\usepackage{soul}
\usepackage{xcolor}
\usepackage{kotex}
\usepackage{hyperref}
\usepackage{cleveref}

\DeclareMathOperator{\Tr}{Tr}

\begin{document}

\title{Exact diffusion coefficients for quantum--classical hybrid walks}

\author{Cook Hyun Kim}
\thanks{These authors contributed equally to this work. Corresponding author: protozerta@kentech.ac.kr}
\email{protozerta@kentech.ac.kr}
\affiliation{CCSS, KI for Grid Modernization, Korea Institute of Energy Technology, 21 Kentech-gil, Naju-si, 58330, Jeollanam-do, Republic of Korea}

\author{Jaesub Park}
\thanks{These authors contributed equally to this work.}
\affiliation{Cambridge Stem Cell Institute, University of Cambridge, Cambridge, CB2 0AW, United Kingdom}
\affiliation{Milner Therapeutics Institute, University of Cambridge, Cambridge, CB2 0AW, United Kingdom}
\affiliation{Cambridge Centre for AI in Medicine, Department of Applied Mathematics and Theoretical Physics, University of Cambridge, Cambridge, CB3 0WA, United Kingdom}

\author{Namshik Han}
\thanks{Corresponding author: nh417@cam.ac.uk}
\email{nh417@cam.ac.uk}
\affiliation{Cambridge Stem Cell Institute, University of Cambridge, Cambridge, CB2 0AW, United Kingdom}
\affiliation{Milner Therapeutics Institute, University of Cambridge, Cambridge, CB2 0AW, United Kingdom}
\affiliation{Cambridge Centre for AI in Medicine, Department of Applied Mathematics and Theoretical Physics, University of Cambridge, Cambridge, CB3 0WA, United Kingdom}
\affiliation{Department of Quantum Information, Institute for Convergence Research and Education in Advanced Technology and Engineering, Yonsei University, Seoul, 03722, Republic of Korea}
\affiliation{Department of Nano Biomedical Engineering (NanoBME), Advanced Science Institute, Yonsei University, Seoul, 03722, Republic of Korea}
\affiliation{Center for Nanomedicine, Institute for Basic Science (IBS), Seoul, 08826, Republic of Korea}

\begin{abstract}
One\hyph{}dimensional quantum--classical hybrid walks are introduced by combining quantum and classical steps under periodic and random protocols. Exact diffusion coefficients are derived for both protocols throughout the diffusive regime. The exact results show how diffusion depends on the average frequency and temporal arrangement of classical steps. As the average frequency approaches zero, both diffusion coefficients diverge inversely with the frequency. At matched average frequencies, the ratio of the diffusion coefficients for the random and periodic protocols approaches two. A finite\hyph{}mode approximation is also developed and tested against the exact diffusion coefficient for the random protocol. The exact results for the one\hyph{}dimensional model provide a foundation for future studies of more general quantum--classical hybrid walks.
\end{abstract}


\maketitle

\section{Introduction}

Quantum walks model coherent transport~\cite{aharonov1993quantum,ambainis2001one,kempe2003quantum,grimmett2004weak}. Applications include quantum algorithms~\cite{shenvi2003quantum,childs2003exponential} and models of biological energy transport~\cite{mohseni2008environment,plenio2008dephasing}. On a line, a Hadamard quantum walk spreads ballistically, with mean squared displacement $\mathrm{MSD}(t)\sim t^2$, whereas an unbiased classical random walk spreads diffusively, with $\mathrm{MSD}(t)\sim t$~\cite{ambainis2001one,kempe2003quantum}. In physical implementations, complete isolation from the environment is difficult, and environmental interactions can suppress quantum coherence~\cite{kendon2003decoherence,kendon2007review}.

Previous studies have examined quantum walks with coin noise~\cite{brun2003decoherent,brun2003transition,abal2008generic}, repeated measurements~\cite{romanelli2005decoherence,romanelli2007measurements,zhang2008limiting,annabestani2010decoherence}, and operations that fluctuate in time or space~\cite{ahlbrecht2011asymptotic,ahlbrecht2012spatiotemporal}. Analytical results include explicit diffusion coefficients for particular coin\hyph{}noise models~\cite{brun2003transition,abal2008generic} and exact moment expressions for walks with coin and position decoherence~\cite{zhang2008limiting,annabestani2010decoherence}. Studies of repeated measurements have related spreading to interval\hyph{}length statistics in the long\hyph{}interval limit~\cite{romanelli2005decoherence,romanelli2007measurements}. The crossover from ballistic spreading to diffusion has also been examined through theoretical analyses and experiments~\cite{kendon2007review,schreiber2011decoherence}. The transport properties depend on how decoherence affects the coin and position.

Decoherence in quantum walks has been extensively studied~\cite{kendon2007review}. Hybrid transport has also been modeled by combining coherent evolution with incoherent transitions~\cite{mohseni2008environment,whitfield2010quantum}. Characterizing diffusion in hybrid walks requires calculating diffusion coefficients for the chosen quantum and classical steps. The present study derives exact diffusion coefficients for discrete\hyph{}time quantum--classical hybrid walks under periodic and random protocols. The model combines Hadamard quantum steps with unbiased classical steps, so that intervals of quantum motion alternate with classical steps. Each classical step first removes coherence between different positions and prepares the coin in an equal mixture of the two directions, while preserving position probabilities. The walker then moves one step left or right with equal probability, independently of the coin state before the classical step. Subsequent quantum steps allow coherent evolution to resume.

The periodic protocol (QRW-P) applies a classical step after a fixed number of quantum steps. The random protocol (QRW-R) applies a classical step independently with a fixed probability at each time step. Both protocols use the same quantum and classical steps. Exact diffusion coefficients are derived for every finite period in the periodic protocol and every nonzero probability of classical steps in the random protocol.

Each interval begins with a classical step and ends immediately before the next classical step. The calculation evaluates the exact displacement variance for every finite interval. For the periodic protocol, the displacement variance gives an exact finite sum for the diffusion coefficient $D_P$. For the random protocol, averaging the displacement variance over a geometric distribution of interval lengths gives a closed\hyph{}form expression for the diffusion coefficient $D_R$. Both coefficients describe long\hyph{}time diffusion, with no approximation in the period or the probability of classical steps.

The exact diffusion coefficients quantify the effect of temporal arrangement at every matched nonzero average frequency of classical steps $\varepsilon$. Both protocols have the same mean interval length. The periodic protocol has fixed interval lengths, whereas the random protocol has fluctuating interval lengths. As $\varepsilon$ approaches zero, both diffusion coefficients diverge as $D\propto\varepsilon^{-1}$, and the ratio $D_R/D_P$ approaches two. During long intervals of quantum motion, ballistic spreading produces a displacement variance that grows quadratically with interval length. Occasional long intervals therefore give the random protocol a larger leading amplitude.

The exact diffusion coefficient for the random protocol provides a benchmark for finite\hyph{}mode approximations. Retaining a finite number of Fourier modes reduces the transfer operator to a finite transfer matrix. The curvature of a transfer\hyph{}matrix eigenvalue gives an approximate diffusion coefficient. Comparison with the exact coefficient quantifies the truncation errors of two finite\hyph{}mode approximations at representative parameter values and in the limit $\varepsilon\to0$. Numerical simulations on a ring independently test the exact diffusion coefficients for both protocols before finite\hyph{}size saturation.

Section~\ref{sec:model} defines the model and protocols. Section~\ref{sec:framework} relates the diffusion coefficient to the curvature of a transfer\hyph{}operator eigenvalue. Sections~\ref{sec:multiplicative} and~\ref{sec:renewal} derive the exact diffusion coefficients for the periodic and random protocols, respectively. Section~\ref{sec:additive} benchmarks the finite\hyph{}mode approximations, and Section~\ref{sec:numerics} compares the exact coefficients with numerical simulations. Section~\ref{sec:discussion} discusses the exact results, the effect of temporal arrangement, and possible extensions.

\section{Quantum and classical dynamics}
\label{sec:model}
The walk takes place on a ring of $N$ nodes and has two possible directions of motion. The Hilbert space is the tensor product of the position and coin spaces,
\begin{equation}
\begin{aligned}
\mathcal H
&= \mathcal H_N \otimes \mathcal H_C, \\[3pt]
\mathcal H_N
&= \operatorname{span}\{ |x\rangle \}_{x=0}^{N-1},
\quad
\mathcal H_C
= \operatorname{span}\{ |+\rangle, |-\rangle \}.
\end{aligned}
\label{eq:hilbert}
\end{equation}
The coin states $|+\rangle$ and $|-\rangle$ represent rightward and leftward motion, with projectors $\Pi_\pm=|\pm\rangle\langle\pm|$. Node indices are taken modulo $N$. The walker is described by the density matrix
\begin{equation}
\rho=\sum_{x,y=0}^{N-1}\sum_{a,b=\pm}\rho_{xa,yb}|x,a\rangle\langle y,b|,
\quad \rho_{xa,yb}=\langle x,a|\rho|y,b\rangle,
\label{eq:density_matrix}
\end{equation}
where $|x,a\rangle=|x\rangle\otimes|a\rangle$, $\rho\ge0$, and $\Tr\rho=1$. The probability of finding the walker at node $x$ is
\begin{equation}
p_x=\Tr_C\langle x|\rho|x\rangle=\sum_{a=\pm}\rho_{xa,xa},
\label{eq:node_probability}
\end{equation}
where $\Tr_C$ denotes the trace over the coin space.

\subsection{Quantum step}
A quantum step applies the Hadamard coin followed by the conditional shift~\cite{aharonov1993quantum,ambainis2001one}. The Hadamard coin mixes the amplitudes of the two directions,
\begin{equation}
H=\frac{1}{\sqrt{2}}\begin{pmatrix}1&1\\1&-1\end{pmatrix}.
\label{eq:coin}
\end{equation}
The conditional shift moves the walker one step to the right for $|+\rangle$ and one step to the left for $|-\rangle$,
\begin{equation}
S=\sum_x\Bigl(|x+1\rangle\langle x|\otimes\Pi_+
+|x-1\rangle\langle x|\otimes\Pi_-\Bigr).
\label{eq:shift}
\end{equation}
The combined operator is $U=S(I_N\otimes H)$, where $I_N$ is the identity on the position space. The quantum step acts on the density matrix as
\begin{equation}
\mathcal U[\rho]=U\rho U^\dagger.
\label{eq:Udef}
\end{equation}
The quantum step preserves quantum coherence and allows interference between different paths.

\subsection{Classical step}
Decoherence acting on the coin and position can be described by maps on the density matrix~\cite{zhang2008limiting,annabestani2010decoherence}. In the present model, a classical step applies the reset operation $\mathcal R$ followed by the conditional shift. The reset operation removes coherence between different positions and prepares the coin in an equal mixture of the two directions,
\begin{equation}
\mathcal R[\rho]=\sum_x p_x|x\rangle\langle x|\otimes\frac{I_C}{2},
\label{eq:reset}
\end{equation}
where $I_C$ is the identity on the coin space. The reset operation $\mathcal R$ preserves the position probabilities and makes the two directions equally likely, independently of the coin state before the classical step.

The conditional shift then moves the walker one step to the right for $|+\rangle$ and one step to the left for $|-\rangle$. The classical step acts on the density matrix as
\begin{equation}
\begin{aligned}
\mathcal T[\rho]&=S\mathcal R[\rho]S^\dagger\\
&=\frac12\sum_x p_x\Bigl(|x+1,+\rangle\langle x+1,+|\\
&\qquad\qquad\qquad+|x-1,-\rangle\langle x-1,-|\Bigr).
\end{aligned}
\label{eq:Tdef}
\end{equation}
The classical step preserves the trace and gives the unbiased classical random-walk update $p'_x=(p_{x-1}+p_{x+1})/2$.

\subsection{Temporal protocols}
Each interval begins with a classical step and ends immediately before the next classical step.

In the periodic protocol (QRW-P), each cycle applies $n_{\mathrm{qw}}$ quantum steps followed by one classical step. The state after one cycle obeys
\begin{equation}
\rho_{t+n_{\mathrm{qw}}+1}
= \mathcal T\bigl[\mathcal U^{n_{\mathrm{qw}}}[\rho_t]\bigr],
\quad t \bmod (n_{\mathrm{qw}}+1)=0.
\label{eq:qrwp}
\end{equation}
The fraction of quantum steps is $r=n_{\mathrm{qw}}/(n_{\mathrm{qw}}+1)$. Classical steps occur at fixed intervals of $n_{\mathrm{qw}}+1$ steps. The fixed interval length defines the coherence timescale,
\begin{equation}
t_P^*=n_{\mathrm{qw}}+1=\frac{1}{1-r}.
\end{equation}

In the random protocol (QRW-R), each time step independently applies $\mathcal U$ with probability $\alpha\in[0,1]$ or $\mathcal T$ with probability $1-\alpha$. The ensemble-averaged state obeys
\begin{equation}
\rho_{t+1}=\alpha\,\mathcal U[\rho_t]+(1-\alpha)\,\mathcal T[\rho_t].
\label{eq:qrwr}
\end{equation}
The mean fraction of quantum steps is $\alpha$. The intervals between classical steps have mean length $1/(1-\alpha)$ for $\alpha<1$. For $0<\alpha<1$, the probability of no classical step during the next $s$ steps is $\alpha^s=\exp(-s/t_R^*)$. The exponential decay defines the coherence timescale,
\begin{equation}
t_R^*=-\frac{1}{\ln\alpha}\sim\frac{1}{1-\alpha}\qquad(\alpha\to1).
\end{equation}
The coherence timescale $t_P^*$ of the periodic protocol equals the fixed interval length. The coherence timescale $t_R^*$ of the random protocol is asymptotically equal to the mean interval length as $\alpha\to1$.

The average frequency of classical steps is $1-r$ in the periodic protocol and $1-\alpha$ in the random protocol. Matching the average frequency $\varepsilon$ of classical steps gives
\begin{equation}
\varepsilon=1-r=1-\alpha,
\quad \alpha=r=\frac{n_{\mathrm{qw}}}{n_{\mathrm{qw}}+1}.
\end{equation}
The comparison therefore uses the discrete values of $r$ allowed by the periodic protocol. At each matched frequency, both protocols have the same mean interval length $1/\varepsilon$. The periodic protocol has fixed interval lengths, whereas the random protocol has fluctuating interval lengths. The symbol $t^*$ denotes the coherence timescale of the relevant protocol. The quantum limit refers to $\varepsilon\to0^+$, whereas the pure quantum walk has no classical steps ($\varepsilon=0$).

\subsection{Displacement and the large-system limit}
The walker starts at node $x=0$ with a symmetric coin state,
\begin{equation}
|\psi_0\rangle=|0\rangle\otimes\frac{|+\rangle+i|-\rangle}{\sqrt2},
\quad \rho_0=|\psi_0\rangle\langle\psi_0|.
\label{eq:init}
\end{equation}
On the ring, the mean squared displacement (MSD) uses the shortest-path distance $d(x,0)$ from the initial node,
\begin{equation}
\mathrm{MSD}(t)=\sum_xd^2(x,0)p_x(t),
\quad \beta(t)=\frac{d\ln\mathrm{MSD}(t)}{d\ln t}.
\label{eq:msd}
\end{equation}
The local exponent $\beta(t)$ distinguishes ballistic spreading, with $\beta=2$, from diffusive spreading, with $\beta=1$.

On the infinite line, each protocol with a nonzero average frequency of classical steps has a diffusion coefficient $D$ defined by the long\hyph{}time growth
\begin{equation}
\langle x^2(t)\rangle=2Dt+O(1).
\label{eq:diffusion_def}
\end{equation}
The exact calculation determines $D$ for every finite $n_{\mathrm{qw}}$ in the periodic protocol and every $0\le\alpha<1$ in the random protocol. Exactness refers to the full parameter dependence of the long\hyph{}time coefficient. On a finite ring, the MSD is bounded. Ring simulations therefore test the infinite-line coefficient during the diffusive regime before finite\hyph{}size saturation.

\section{Spectral framework for the diffusion coefficient}
\label{sec:framework}
\subsection{Fourier representation}
The Fourier representation assigns one variable to each position index of the density matrix~\cite{zhang2008limiting,annabestani2010decoherence},
\begin{equation}
\begin{aligned}
\tilde\rho_t(k,q)&=\sum_{x,y}e^{ikx-iqy}\rho_t(x,y),\\[2pt]
\rho_t(x,y)&=\langle x|\rho_t|y\rangle.
\end{aligned}
\label{eq:fourier_main}
\end{equation}
Each $\rho_t(x,y)$ is a $2\times2$ matrix in the coin space. On a ring of $N$ nodes, $k$ and $q$ take the values $2\pi j/N$ with $j=0,\ldots,N-1$. The single-step expressions below omit the time index.

The quantum step $U=S(I_N\otimes H)$ has the momentum-space representation
\begin{equation}
V(k)=E(k)H,
\quad E(k)=\begin{pmatrix}e^{ik}&0\\0&e^{-ik}\end{pmatrix}.
\label{eq:QLQR}
\end{equation}
The update $\mathcal U[\rho]=U\rho U^\dagger$ therefore becomes
\begin{equation}
\tilde{\mathcal U}[\tilde\rho](k,q)=V(k)\tilde\rho(k,q)V^\dagger(q).
\label{eq:U_momentum}
\end{equation}

The classical step depends only on the position probabilities $p_x=\Tr_C\rho(x,x)$. In position space,
\begin{equation}
\mathcal T[\rho]=\frac12\sum_x p_x\Bigl(|x+1\rangle\langle x+1|\otimes\Pi_++|x-1\rangle\langle x-1|\otimes\Pi_-\Bigr).
\end{equation}
Define $u=k-q$ and $w=k+q$. Fourier transformation of the two position projectors gives the factors $e^{iu(x+1)}$ and $e^{iu(x-1)}$, respectively. Since $e^{iu}\Pi_++e^{-iu}\Pi_-=E(u)$, the transformed classical step is
\begin{equation}
\tilde{\mathcal T}[\tilde\rho](u)=\frac{E(u)}2\sum_x e^{iux}p_x.
\end{equation}

The momentum average at fixed $u$ connects $\sum_x e^{iux}p_x$ to $\tilde\rho$:
\begin{equation}
\langle F\rangle_q=\frac1N\sum_q F(q+u,q).
\label{eq:average}
\end{equation}
Substituting Eq.~\eqref{eq:fourier_main} and using $N^{-1}\sum_q e^{iq(x-y)}=\delta_{xy}$ gives
\begin{equation}
\begin{aligned}
\langle\Tr_C\tilde\rho\rangle_q
&=\sum_{x,y}e^{iux}\Tr_C\rho(x,y)\frac1N\sum_q e^{iq(x-y)}\\
&=\sum_x e^{iux}\Tr_C\rho(x,x)\\
&=\sum_x e^{iux}p_x.
\end{aligned}
\end{equation}
The momentum average selects the position-diagonal elements, and the coin trace gives the position probabilities. The classical step therefore becomes
\begin{equation}
\tilde{\mathcal T}[\tilde\rho](u)=\frac{E(u)}2\langle\Tr_C\tilde\rho\rangle_q.
\label{eq:Delta_momentum}
\end{equation}
In the infinite-line limit, $N^{-1}\sum_q$ becomes the normalized integral $\int_{-\pi}^{\pi}dq/(2\pi)$.

\subsection{Diffusion from eigenvalue curvature}
Eigenvalue perturbation provides a method for calculating the long\hyph{}time displacement moments of quantum walks~\cite{ahlbrecht2011asymptotic,ahlbrecht2012spatiotemporal}. The transfer operators for the periodic and random protocols are
\begin{equation}
\mathcal W_P=\tilde{\mathcal T}\circ\tilde{\mathcal U}^{n_{\mathrm{qw}}},
\quad \mathcal W_R=\alpha\tilde{\mathcal U}+(1-\alpha)\tilde{\mathcal T}.
\label{eq:transfer_main}
\end{equation}
The operator $\mathcal W_P$ advances one cycle of $\tau=n_{\mathrm{qw}}+1$ steps, whereas $\mathcal W_R$ advances one step ($\tau=1$). Neither operator mixes different values of $u$, so the eigenvalue problem separates at each fixed $u$.

Taking the coin trace and averaging over $q$ at fixed $u$ gives the displacement characteristic function,
\begin{equation}
\chi_t(u)=\langle\Tr_C\tilde\rho_t\rangle_q=\sum_xe^{iux}p_x(t).
\label{eq:charfn}
\end{equation}
Derivatives of $\chi_t(u)$ at $u=0$ give displacement moments.

Let $\lambda(u)$ denote the dominant eigenvalue near $u=0$, with $\lambda(0)=1$. Under the assumption that $\lambda(u)$ is simple, isolated, and larger in absolute value than the remaining spectrum, the eigenvalue determines the long\hyph{}time evolution of $\chi_t(u)$:
\begin{equation}
\ln\chi_t(u)=\frac{t}{\tau}\ln\lambda(u)+O(1).
\label{eq:cgf}
\end{equation}
Left--right symmetry gives zero mean displacement, so the second derivative of $\ln\lambda(u)$ determines the diffusion coefficient,
\begin{equation}
D=-\frac1{2\tau}\left.\frac{d^2}{du^2}\ln\lambda(u)\right|_{u=0}.
\label{eq:msd_eigen_main}
\end{equation}
Thus, if $\lambda(u)=1+\gamma u^2+O(u^4)$, then $D=-\gamma/\tau$.

The expansion around $u=0$ extracts the curvature that determines $D$ while retaining the full dependence on the protocol parameters. For the periodic protocol, extracting the constant Laurent coefficient evaluates the eigenvalue curvature exactly. For the random protocol, averaging the exact displacement variance over the interval\hyph{}length distribution gives the diffusion coefficient. Section~\ref{sec:additive} uses the same curvature formula to benchmark a finite\hyph{}mode approximation.

\section{Exact diffusion coefficient for the periodic protocol}
\label{sec:multiplicative}
The periodic protocol admits an exact diffusion coefficient for every finite period. Define $n\equiv n_{\mathrm{qw}}$, so each period contains $n+1$ steps. The derivation first evaluates the transfer\hyph{}operator eigenvalue curvature through a Laurent recurrence, then expresses the result as a finite sum.

\subsection{Calculating the transfer-operator eigenvalue}
At fixed $u$, the classical step gives $c(u)E(u)/2$, where $c(u)$ depends on the input state. Applying $n$ quantum steps gives
\begin{equation}
\tilde{\mathcal U}^{n}\left[c(u)\frac{E(u)}2\right]
=\frac{c(u)}2V(k)^nE(u)V^\dagger(q)^n.
\end{equation}
The next classical step takes the coin trace, averages over $q$, and multiplies by $E(u)/2$. The resulting state is $\lambda_P(u)c(u)E(u)/2$, where the nonzero eigenvalue of $\mathcal W_P$ is
\begin{equation}
\lambda_P(u)=\frac12\left\langle\Tr_C\left[V(k)^nE(u)V^\dagger(q)^n\right]\right\rangle_q.
\label{eq:cycle_eigenvalue}
\end{equation}

The matrix powers in Eq.~\eqref{eq:cycle_eigenvalue} are evaluated from the eigenvalues of $V(k)$, following the Fourier analysis of the Hadamard walk~\cite{aharonov1993quantum,ambainis2001one}. The trace and determinant of $V(k)$ are
\begin{equation}
\begin{aligned}
\Tr_C V(k)&=\sqrt2\,i\sin k,\\[2pt]
\det V(k)&=-1.
\end{aligned}
\label{eq:trdet}
\end{equation}
The eigenvalues are $e^{i\omega_k}$ and $-e^{-i\omega_k}$, where
\begin{equation}
\sqrt2\sin\omega_k=\sin k.
\label{eq:dispersion}
\end{equation}
Evaluating the matrix powers with the two eigenvalues and expanding around $u=0$ at fixed $w=k+q$ gives
\begin{equation}
\frac12\Tr_C\left[V(k)^nE(u)V^\dagger(q)^n\right]=1+r_2(z)u^2+O(u^4).
\end{equation}
Here $z=e^{iw}$, and $r_2(z)$ is the coefficient of $u^2$. For finite $n$, $r_2(z)$ is a Laurent polynomial.

The classical step includes the momentum average in Eq.~\eqref{eq:Delta_momentum}. In the infinite-line limit, the average of a $2\pi$-periodic function of $w=2q+u$ over $q$ equals the average over one period in $w$. Each power $z^m=e^{imw}$ therefore has the average
\begin{equation}
\int_{-\pi}^{\pi}\frac{dw}{2\pi}e^{imw}
=\begin{cases}
1,&m=0,\\
0,&m\ne0.
\end{cases}
\end{equation}
The momentum average therefore removes all nonzero powers of $z$ and retains only the constant Laurent coefficient,
\begin{equation}
[r_2]_{z^0}=\int_{-\pi}^{\pi}\frac{dw}{2\pi}r_2(e^{iw}).
\label{eq:bracketdef}
\end{equation}
The eigenvalue of $\mathcal W_P$ is then
\begin{equation}
\lambda_P(u)=1+[r_2]_{z^0}u^2+O(u^4).
\label{eq:lambda_z0}
\end{equation}

\subsection{Calculating the quadratic coefficient}
The quadratic coefficient is
\begin{equation}
r_2(z)=\frac{c_0+c_1n+c_2n^2}{2P(z)^2}
+\frac{z^2\sigma_{n+1}(z)}{4^nP(z)^2},
\label{eq:r2}
\end{equation}
where
\begin{equation}
\begin{aligned}
P(z)&=1+6z+z^2,\\[2pt]
c_0&=-(1+8z+30z^2+8z^3+z^4),\\[2pt]
c_1&=-2(1+z)^2P(z),\\[2pt]
c_2&=-(1+z)^2P(z).
\end{aligned}
\label{eq:cs}
\end{equation}
The term $\sigma_{n+1}(z)$ is defined by
\begin{equation}
\begin{aligned}
\sigma_p(z)&=C_+(z)^p+C_-(z)^p,\\[2pt]
C_\pm(z)&=\frac{-(1+z)^2\pm(z-1)\sqrt{P(z)}}{z}.
\end{aligned}
\label{eq:Cpm}
\end{equation}
The square roots in $C_\pm$ cancel in the sum and product:
\begin{equation}
C_++C_-=-\frac{2(1+z)^2}{z},
\quad C_+C_-=16.
\label{eq:sympoly}
\end{equation}
Using $\sigma_{p+1}=(C_++C_-)\sigma_p-C_+C_-\sigma_{p-1}$ gives
\begin{equation}
\begin{aligned}
\sigma_{p+1}&=-\frac{2(1+z)^2}{z}\sigma_p-16\sigma_{p-1},\\[2pt]
\sigma_0&=2,\quad \sigma_1=-\frac{2(1+z)^2}{z}.
\end{aligned}
\label{eq:recurrence}
\end{equation}
Equation~\eqref{eq:recurrence} gives $\sigma_{n+1}(z)$ as a Laurent polynomial without evaluating square roots.

\subsection{Calculating the diffusion coefficient}
The constant Laurent coefficient of $r_2(z)$ determines the diffusion coefficient. Writing $P=P(z)$ and using $c_1=2c_2$ and $c_0=c_2-16z^2$ gives
\begin{equation}
\frac{c_0+c_1n+c_2n^2}{2P^2}=-\frac{(n+1)^2(1+z)^2}{2P}-\frac{8z^2}{P^2}.
\label{eq:Psplit}
\end{equation}
Expanding each term about $z=0$ gives the constant coefficients $-(n+1)^2/2$ and zero, respectively. Thus,
\begin{equation}
[r_2]_{z^0}=-\frac{(n+1)^2}{2}+A_n,
\quad A_n=\left[\frac{z^2\sigma_{n+1}(z)}{4^nP(z)^2}\right]_{z^0}.
\label{eq:r2z0}
\end{equation}

Expansions of $\sigma_p(z)$ and $P(z)^{-2}$ about $z=0$ give a finite recurrence calculation of $A_n$:
\begin{equation}
\sigma_p(z)=\sum_j s_{p,j}z^j,
\quad P(z)^{-2}=\sum_{j\ge0}b_jz^j.
\label{eq:coefficient_arrays}
\end{equation}
Equation~\eqref{eq:recurrence} determines $s_{p,j}$. Matching powers of $z$ in $P(z)^2\sum_{j\ge0}b_jz^j=1$ determines $b_j$. Expanding $P(z)^2$ gives
\begin{equation}
P(z)^2=1+12z+38z^2+12z^3+z^4.
\end{equation}
Matching coefficients then gives
\begin{equation}
\begin{aligned}
b_j&=-12b_{j-1}-38b_{j-2}-12b_{j-3}-b_{j-4},\quad j\ge1,\\[2pt]
b_0&=1,\quad b_j=0\quad(j<0).
\end{aligned}
\label{eq:b_recurrence}
\end{equation}
In the product $z^2\sigma_{n+1}(z)P(z)^{-2}$, the term $b_jz^j$ contributes to the constant coefficient only when paired with $s_{n+1,-j-2}z^{-j-2}$. The lowest power in $\sigma_{n+1}(z)$ is $-(n+1)$, so $j$ ranges from $0$ to $n-1$. Therefore,
\begin{equation}
A_n=4^{-n}\sum_{j=0}^{n-1}b_j s_{n+1,-j-2},\quad n\ge1.
\label{eq:A_convolution}
\end{equation}
Equation~\eqref{eq:A_convolution} gives
\begin{equation}
A_n=1,\ 3,\ \frac{11}{2},\ \frac{17}{2},\ \frac{99}{8},\ \frac{137}{8},\ldots,
\quad n=1,2,\ldots.
\label{eq:Avalues}
\end{equation}

Using Eq.~\eqref{eq:r2z0} in the diffusion formula, Eq.~\eqref{eq:msd_eigen_main}, with $\tau=n+1$ gives
\begin{equation}
D_P(n)=-\frac{[r_2]_{z^0}}{n+1}=\frac{n+1}{2}-\frac{A_n}{n+1}.
\label{eq:DP}
\end{equation}
The first five values are
\begin{equation}
D_P(n)=\frac12,\ \frac12,\ \frac58,\ \frac45,\ \frac{15}{16},
\quad n=1,\ldots,5.
\label{eq:DPvalues}
\end{equation}
For $n=0$, each cycle consists of one classical step, so $D_P(0)=1/2$. Equation~\eqref{eq:DP} includes the classical limit with $A_0=0$.

\subsection{Generating function and exact finite sum}
The generating function of $A_n$ gives an explicit finite sum for $D_P(n)$ and evaluates the average over interval lengths needed for $D_R(\alpha)$.

Choose the square-root branch with $\sqrt{P(0)}=1$. Near $z=0$, $C_+\sim-2/z$ and $C_-\sim-8z$. The term $z^2C_-^{n+1}/(4^nP^2)$ contains only positive powers of $z$ and has no constant coefficient. Thus,
\begin{equation}
A_n=\left[\frac{z^2C_+^{n+1}}{4^nP^2}\right]_{z^0}.
\label{eq:onlyCp}
\end{equation}
To evaluate the constant coefficient, set $C_+=-4\xi$. The product $C_+C_-=16$ gives $C_-=-4/\xi$, and the sum in Eq.~\eqref{eq:sympoly} gives
\begin{equation}
z+z^{-1}=2(\xi+\xi^{-1})-2.
\label{eq:joukowski}
\end{equation}
As $z\to0$, $\xi\sim1/(2z)$ and therefore $\xi\to\infty$. The change from $z$ to $\xi$ gives
\begin{equation}
P=\frac{2z(\xi+1)^2}{\xi},
\quad \frac{dz}{z}=-\frac{\xi+1}{\xi\sqrt{\xi^2+1}}\,d\xi,
\label{eq:jacobian}
\end{equation}
where the square-root branch satisfies $\sqrt{\xi^2+1}\sim\xi$ as $\xi\to\infty$.

Writing the constant coefficient as a contour integral and changing variables from $z$ to $\xi$ yields
\begin{equation}
A_n=(-1)^{n+1}\left[\frac{\xi^{n+2}}{(\xi+1)^3\sqrt{\xi^2+1}}\right]_{\xi^{-1}}.
\label{eq:Aw}
\end{equation}
The brackets select the coefficient of $\xi^{-1}$ in the expansion at infinity. Setting $x=1/\xi$ gives
\begin{equation}
\frac{\xi^{n+2}}{(\xi+1)^3\sqrt{\xi^2+1}}
=x^{2-n}\frac1{(1+x)^3\sqrt{1+x^2}}.
\end{equation}
The coefficient of $\xi^{-1}$ equals the coefficient of $x^1$. Factoring out $x^{2-n}$ reduces the extraction to the coefficient of $x^{n-1}$ in the remaining function. Thus,
\begin{equation}
A_n=(-1)^{n+1}\left[\frac1{(1+x)^3\sqrt{1+x^2}}\right]_{x^{n-1}}.
\label{eq:Ax}
\end{equation}
Replacing $x$ by $-s$ cancels the factor $(-1)^{n+1}$, so $A_n$ is the coefficient of $s^{n-1}$ in $(1-s)^{-3}(1+s^2)^{-1/2}$. The generating function is therefore
\begin{equation}
\Phi(s)\equiv\sum_{n\ge1}A_ns^{n-1}=\frac1{(1-s)^3\sqrt{1+s^2}}.
\label{eq:genfunc}
\end{equation}

Expanding the two factors in Eq.~\eqref{eq:genfunc} and collecting the coefficient of $s^{n-1}$ gives
\begin{equation}
A_n=\sum_{j=0}^{\lfloor(n-1)/2\rfloor}\frac{(-1)^j}{4^j}\binom{2j}{j}\binom{n+1-2j}{2},
\quad n\ge1.
\label{eq:A_finite_sum}
\end{equation}
Equations~\eqref{eq:DP} and~\eqref{eq:A_finite_sum} give an explicit finite sum for the diffusion coefficient at every finite $n\ge1$. Setting $A_0=0$ includes $n=0$. The finite sum evaluates the momentum average exactly and retains all finite\hyph{}interval contributions.

\subsection{Quantum limit}

The behavior of $\Phi(s)$ near $s=1$ determines the large-$n$ limit. The factor $(1-s)^{-3}$ produces quadratic growth in $A_n$, while $(1+s^2)^{-1/2}$ approaches $1/\sqrt2$. The branch points at $s=\pm i$ contribute lower-order oscillatory terms. Hence,
\begin{equation}
A_n=\frac{n(n+1)}{2\sqrt2}+O(n).
\label{eq:bracketasym}
\end{equation}
Substituting Eq.~\eqref{eq:bracketasym} into Eq.~\eqref{eq:DP} and using $1-r=1/(n+1)$ gives
\begin{equation}
D_P(n)=\frac{2-\sqrt2}{4}\,n+O(1)\sim\frac{2-\sqrt2}{4(1-r)}.
\label{eq:DPasym}
\end{equation}
Thus, the diffusion coefficient diverges inversely with the average frequency of classical steps. For $n=100$, $800$, and $1600$, the values $D_P/n=0.14794$, $0.14663$, and $0.14654$, respectively, approach the quantum\hyph{}limit amplitude $(2-\sqrt2)/4\simeq0.146447$.

\section{Exact diffusion coefficient for the random protocol}
\label{sec:renewal}

The random protocol admits a closed\hyph{}form diffusion coefficient for every $0\le\alpha<1$. The derivation averages the exact displacement variance over a geometric distribution of interval lengths and evaluates the resulting infinite sum with the generating function in Eq.~\eqref{eq:genfunc}.

\subsection{Intervals between classical steps}

An interval contains one classical step followed by $n$ quantum steps and has length $\ell=n+1$. The initial classical step applies $\mathcal R$ followed by the conditional shift. Since $H(I_C/2)H^\dagger=I_C/2$, the initial conditional shift has the same action as a quantum step on the state prepared by the reset operation. The displacement variance $V_\ell$ therefore equals the displacement variance of a pure quantum walk of $\ell$ steps starting from a localized position and coin state $I_C/2$. The periodic protocol with $n_{\mathrm{qw}}=n$ gives
\begin{equation}
V_{n+1}=2(n+1)D_P(n).
\label{eq:segment_variance}
\end{equation}

In the random protocol, each step is independently quantum with probability $\alpha$ and classical with probability $1-\alpha$. The number $n$ of quantum steps between successive classical steps therefore has the geometric distribution
\begin{equation}
p_n=(1-\alpha)\alpha^n,\quad n\ge0.
\label{eq:geometric}
\end{equation}
The mean interval length is
\begin{equation}
\sum_{n\ge0}p_n(n+1)=\frac{1}{1-\alpha}.
\end{equation}

The reset operation $\mathcal R$ removes quantum coherence and prepares the coin in the state $I_C/2$. Translation invariance makes the displacement distribution within an interval independent of the starting node. For a fixed interval length, the displacement distribution is also independent of earlier intervals and the starting time. Successive interval displacements are independent and have zero mean, so the displacement variances add.

\subsection{Exact averaging and closed-form diffusion coefficient}

Studies of repeated measurements calculate spreading by adding the displacement variances accumulated between measurements~\cite{romanelli2005decoherence,romanelli2007measurements}. For the present hybrid model, dividing the mean displacement variance per interval by twice the mean interval length gives
\begin{equation}
\begin{aligned}
D_R(\alpha)
&=\frac{\sum_{n\ge0}p_nV_{n+1}}
       {2\sum_{n\ge0}p_n(n+1)}\\
&=(1-\alpha)^2\sum_{n\ge0}\alpha^n(n+1)D_P(n).
\end{aligned}
\label{eq:renewal_transform}
\end{equation}
For fixed $\alpha<1$, the mean displacement variance contributed by the incomplete final interval remains bounded as the observation time increases. The incomplete interval therefore leaves the long\hyph{}time diffusion coefficient unchanged. The long\hyph{}time limit precedes the quantum limit $\alpha\to1$.

Substituting Eq.~\eqref{eq:DP} into Eq.~\eqref{eq:renewal_transform} gives
\begin{equation}
D_R(\alpha)=(1-\alpha)^2\left[\frac12\sum_{n\ge0}(n+1)^2\alpha^n-\sum_{n\ge0}A_n\alpha^n\right].
\label{eq:DRseries}
\end{equation}
Both sums converge absolutely for $0\le\alpha<1$. Evaluating the first sum and using Eq.~\eqref{eq:genfunc} for the second gives
\begin{equation}
\begin{aligned}
\sum_{n\ge0}(n+1)^2\alpha^n
&=\frac{1+\alpha}{(1-\alpha)^3},\\[2pt]
\sum_{n\ge0}A_n\alpha^n
&=\alpha\Phi(\alpha)=\frac{\alpha}{(1-\alpha)^3\sqrt{1+\alpha^2}}.
\end{aligned}
\end{equation}
The exact diffusion coefficient is therefore
\begin{equation}
D_R(\alpha)
=\frac{1+\alpha-2\alpha/\sqrt{1+\alpha^2}}
       {2(1-\alpha)}.
\label{eq:DRexact}
\end{equation}

Equation~\eqref{eq:DRexact} completes the interval average for every $0\le\alpha<1$. The closed\hyph{}form expression retains contributions from all finite interval lengths and requires no expansion in $1-\alpha$. For $\alpha=0$, every step is classical, so $D_R(0)=1/2$.

\subsection{Quantum limit}

Expanding the exact diffusion coefficient as $\alpha\to1$ gives
\begin{equation}
D_R(\alpha)\sim\frac{C_R}{1-\alpha},
\quad C_R=1-\frac{1}{\sqrt2}=\frac{2-\sqrt2}{2}.
\label{eq:DRexactpole}
\end{equation}
The diffusion coefficient thus diverges inversely with the average frequency of classical steps, as in the periodic protocol. Comparing with Eq.~\eqref{eq:DPasym} gives
\begin{equation}
C_R=2C_P,\quad C_P=\frac{2-\sqrt2}{4}.
\label{eq:amplitude_ratio}
\end{equation}
At the same average frequency $\varepsilon=1-\alpha=1-r$, the ratio $D_R/D_P$ approaches $2$ as $\varepsilon\to0$. Equations~\eqref{eq:DP}, \eqref{eq:A_finite_sum}, and~\eqref{eq:DRexact} also give the exact comparison at every matched nonzero frequency. Section~\ref{sec:discussion} explains the amplitude difference through the interval\hyph{}length statistics.

\section{Benchmarking finite-mode approximations}
\label{sec:additive}

The exact diffusion coefficient $D_R(\alpha)$ in Eq.~\eqref{eq:DRexact} provides a benchmark for finite\hyph{}mode approximations. Retaining the Fourier modes $|m|\le L$ in the transfer operator $\mathcal W_R$ gives the finite transfer matrix $\mathcal W_R^{(L)}$. The curvature of the eigenvalue $\lambda_L(u)$ with $\lambda_L(0)=1$ gives the approximate diffusion coefficient $D_R^{(L)}(\alpha)$. Comparing the explicit expressions for $L=1$ and $L=2$ with the exact coefficient quantifies the improvement from retaining additional modes.

\subsection{Pauli representation and Fourier modes}

The Fourier-transformed density matrix has the Pauli expansion~\cite{brun2003decoherent,abal2008generic,annabestani2010decoherence}
\begin{equation}
\begin{aligned}
\tilde\rho
&=\frac12(aI_C+b\sigma_z+c\sigma_x+d\sigma_y),\\[2pt]
\mathbf v&=(a,b,c,d)^{\mathsf T}.
\end{aligned}
\label{eq:pauli}
\end{equation}
The coefficients are generally complex for $k\ne q$. With $u=k-q$, $w=k+q$, and $z=e^{iw}$, the Fourier expansion at fixed $u$ is
\begin{equation}
\mathbf v(u,w)=\sum_m\mathbf v_m(u)z^m.
\end{equation}
Each Fourier mode has four Pauli components.

The matrix $Q(u,w)=Q_0+zQ_++z^{-1}Q_-$ represents the quantum step. The matrix $T_0$, acting only on $m=0$, represents the classical step. The matrices are
\begin{equation}
\begin{aligned}
Q_0&=\begin{pmatrix}
\cos u&0&i\sin u&0\\
i\sin u&0&\cos u&0\\
0&0&0&0\\
0&0&0&0
\end{pmatrix},\\[4pt]
Q_\pm&=\frac12\begin{pmatrix}
0&0&0&0\\
0&0&0&0\\
0&1&0&\pm i\\
0&\pm i&0&-1
\end{pmatrix},
\quad T_0=\begin{pmatrix}
\cos u&0&0&0\\
i\sin u&0&0&0\\
0&0&0&0\\
0&0&0&0
\end{pmatrix}.
\end{aligned}
\label{eq:transfer_blocks}
\end{equation}
The factors $z^{\pm1}$ couple neighboring Fourier modes, while the classical step retains only $m=0$. The transfer operator therefore has the blocks
\begin{equation}
\begin{aligned}
(\mathcal W_R)_{mm'}
&=\alpha Q_0\delta_{mm'}+\alpha Q_+\delta_{m,m'+1}\\
&\quad+\alpha Q_-\delta_{m,m'-1}+(1-\alpha)T_0\delta_{m0}\delta_{m'0}.
\end{aligned}
\label{eq:block}
\end{equation}

\subsection{Truncation and diffusion coefficient}

Each classical step removes all Fourier modes with $m\ne0$, motivating a truncation near $m=0$. The truncation retains $|m|\le L$ and omits couplings to modes with $|m|>L$. The resulting transfer matrix $\mathcal W_R^{(L)}$ has dimension $4(2L+1)$.

For each Fourier cutoff $L$, the eigenvalue with $\lambda_L(0)=1$ has the expansion
\begin{equation}
\begin{aligned}
\det[\mathcal W_R^{(L)}(u)-\lambda_L(u)\mathbf1]&=0,\\[2pt]
\lambda_L(u)&=1+\gamma_L(\alpha)u^2+O(u^4).
\end{aligned}
\label{eq:characteristic}
\end{equation}
For $0\le\alpha<1$, the eigenvalue at $u=0$ is simple and isolated, allowing nondegenerate perturbation theory. Equation~\eqref{eq:msd_eigen_main} gives $D_R^{(L)}=-\gamma_L$.

For $L=1$, the transfer matrix has dimension $12$ and gives
\begin{equation}
D_R^{(1)}(\alpha)
=\frac{2+\alpha^3+\alpha^4}
       {2(2-\alpha^3-\alpha^4)}.
\label{eq:DR12}
\end{equation}
For $L=2$, the transfer matrix has dimension $20$ and gives
\begin{equation}
D_R^{(2)}(\alpha)=\frac{4+4\alpha+3\alpha^2+3\alpha^3+4\alpha^4+4\alpha^5+2\alpha^6}{2(4+4\alpha+3\alpha^2-\alpha^3-4\alpha^4-4\alpha^5-2\alpha^6)}.
\label{eq:DR20}
\end{equation}
Both approximations recover the classical\hyph{}limit diffusion coefficient, $D_R^{(L)}(0)=D_R(0)=1/2$.

\subsection{Comparison with the exact diffusion coefficient}

The relative error is
\begin{equation}
e_L(\alpha)
=\left|\frac{D_R^{(L)}(\alpha)}{D_R(\alpha)}-1\right|.
\label{eq:relative_error}
\end{equation}
Table~\ref{tab:approx} lists the exact diffusion coefficient and the relative errors for $L=1$ and $L=2$. Increasing the cutoff reduces the relative error at every listed value of $\alpha$. At $\alpha=0.5$, the error decreases from about $0.351\%$ to $0.00746\%$; at $\alpha=0.99$, the error decreases from about $2.42\%$ to $0.0708\%$.

\begin{table}[t]
\caption{Exact diffusion coefficient and relative errors of the finite\hyph{}mode approximations for the random protocol. Relative errors are reported as $100e_L(\alpha)$. All values are calculated from Eqs.~\eqref{eq:DRexact}, \eqref{eq:DR12}, and \eqref{eq:DR20}.}
\label{tab:approx}
\centering
\begin{tabular}{cccc}
\toprule
$\alpha$ & Exact $D_R$ & $L=1$ error (\%) & $L=2$ error (\%)\\
\midrule
0.10 & 0.50055142 & 0.000224 & $4.56\times10^{-7}$ \\
0.50 & 0.60557281 & 0.350830 & 0.007459 \\
0.80 & 1.37652476 & 1.592805 & 0.045120 \\
0.95 & 5.72501076 & 2.264739 & 0.066241 \\
0.99 & 29.14554021 & 2.415705 & 0.070763 \\
\bottomrule
\end{tabular}
\end{table}

\subsection{Quantum limit}

The denominators in Eqs.~\eqref{eq:DR12} and~\eqref{eq:DR20} have simple zeros at $\alpha=1$. Expanding near the quantum limit gives
\begin{equation}
D_R^{(L)}(\alpha)\sim\frac{C_R^{(L)}}{1-\alpha},
\quad C_R^{(1)}=\frac27,
\quad C_R^{(2)}=\frac{12}{41}.
\label{eq:DRpole}
\end{equation}
Both approximations reproduce the inverse-frequency divergence of the exact diffusion coefficient. The relative error approaches the relative error in the quantum\hyph{}limit amplitude:
\begin{equation}
\lim_{\alpha\to1^-}e_L(\alpha)
=\left|\frac{C_R^{(L)}}{C_R}-1\right|.
\label{eq:amplitude_error}
\end{equation}
Using the exact quantum\hyph{}limit amplitude $C_R=1-1/\sqrt2$, the limiting relative error is approximately $2.45\%$ for $L=1$ and $0.072\%$ for $L=2$.

The finite\hyph{}mode calculation obtains diffusion coefficients from the transfer matrix without an exact expression for the displacement variance. Comparison with Eq.~\eqref{eq:DRexact} gives relative errors below $0.1\%$ for $L=2$ at the values in Table~\ref{tab:approx} and in the quantum limit. Modifying the transfer matrix extends the approach to other coin operations or classical steps. Each extension requires checking the spectral conditions and evaluating the truncation error.

\section{Numerical validation}
\label{sec:numerics}

Numerical simulations test the exact diffusion coefficients on a ring of $N=2501$ nodes for up to $T=5\times10^4$ time steps. The calculation uses double-precision complex arithmetic and applies Eqs.~\eqref{eq:qrwp} and~\eqref{eq:qrwr} directly to the density matrix. The updates average over the random protocol without sampling individual trajectories and preserve normalization up to round-off error.

The extraction of $D$ has three steps. First, divide each MSD time series into consecutive 200-step windows. Second, fit the local log--log slope $\beta$ in each window and retain windows with $|\beta-1|<0.05$. Third, fit $\mathrm{MSD}(t)=2Dt+c$ in each retained window and average the fitted diffusion coefficients. Each estimate requires at least three retained windows.

\begin{figure}[t]
\centering
\includegraphics[width=\columnwidth]{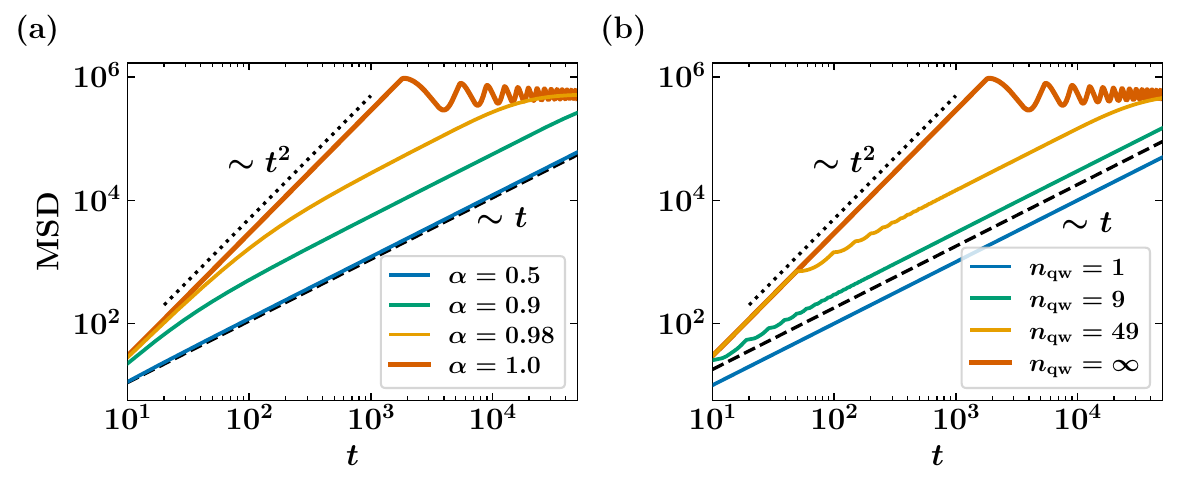}
\caption{Mean squared displacement on a ring of $N=2501$ nodes over $T=5\times10^4$ time steps, shown on logarithmic axes. (a)~Random protocol (QRW-R) at $\alpha=0.5$, $0.9$, $0.98$, and the pure quantum walk ($\alpha=1$). (b)~Periodic protocol (QRW-P) at $n_{\mathrm{qw}}=1$, $9$, $49$, and the pure quantum walk (no classical steps). Dashed and dotted lines indicate slopes $1$ and $2$, respectively. The hybrid walks develop diffusive regimes with $\beta\simeq1$. The pure quantum walks spread ballistically with $\beta\simeq2$ before finite\hyph{}size effects appear near $t\simeq2\times10^3$.}
\label{fig:msd_ring}
\end{figure}

Figure~\ref{fig:msd_ring} shows the crossover from ballistic to diffusive spreading in the hybrid walks. The pure quantum walk reaches the ring scale near $t\simeq2\times10^3$. The estimates of the diffusion coefficients for the hybrid walks use diffusive windows before finite\hyph{}size saturation.

\begin{figure}[t]
\centering
\includegraphics[width=\columnwidth]{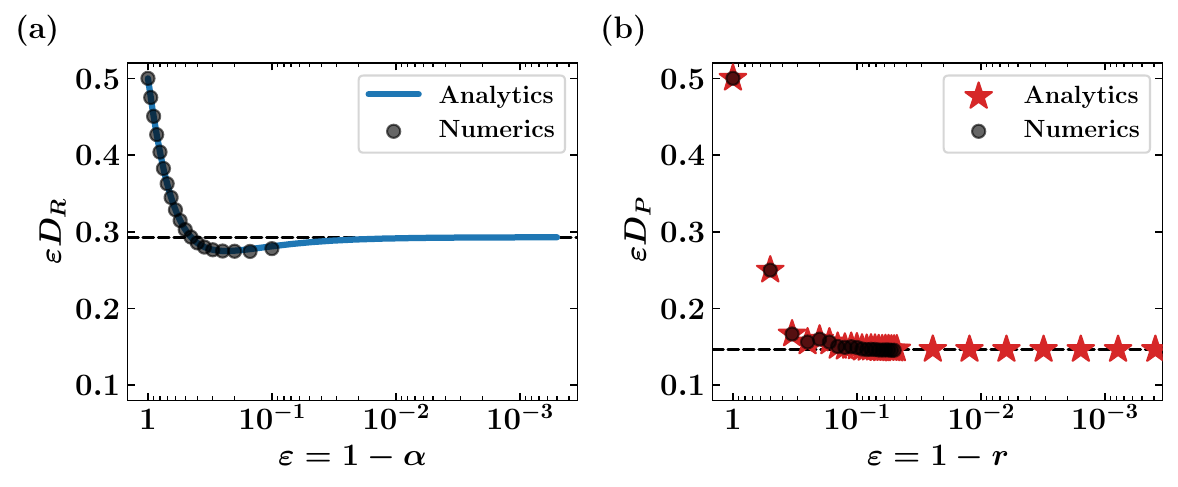}
\caption{Rescaled diffusion coefficient $\varepsilon D$ versus the average frequency of classical steps: (a)~$\varepsilon=1-\alpha$ for the random protocol (QRW-R); (b)~$\varepsilon=1-r$ for the periodic protocol (QRW-P). The logarithmic horizontal axes decrease toward the quantum limit. Black circles show diffusion coefficients extracted from ring simulations with $N=2501$, covering $\alpha\le0.95$ in (a) and $n_{\mathrm{qw}}\le19$ in (b). The blue solid line in (a) shows the exact coefficient $D_R$ from Eq.~\eqref{eq:DRexact}. Red stars in (b) show the exact coefficient $D_P$ calculated from the Laurent recurrence at selected integer values of $n_{\mathrm{qw}}$ up to $2560$. Dashed lines indicate the exact quantum\hyph{}limit amplitudes $C_R=(2-\sqrt2)/2$ and $C_P=(2-\sqrt2)/4$. Both protocols give $\varepsilon D=1/2$ in the classical limit $\varepsilon=1$.}
\label{fig:D_ring}
\end{figure}

Figure~\ref{fig:D_ring} compares the diffusion coefficients estimated from simulations with the exact diffusion coefficients $D_R$ for the random protocol and $D_P$ for the periodic protocol. The simulations test $\alpha\le0.95$ in the random protocol and $n_{\mathrm{qw}}\le19$ in the periodic protocol. The exact formulas cover the full parameter ranges, including smaller average frequencies of classical steps where the crossover to diffusion takes longer.

\section{Discussion and conclusions}
\label{sec:discussion}

\subsection{Exact diffusion coefficients}

The main result is the exact calculation of diffusion coefficients for periodic and random protocols that use the same quantum and classical steps. Equations~\eqref{eq:DP} and~\eqref{eq:A_finite_sum} give an exact finite sum for $D_P(n)$ at every finite $n\ge0$, with $A_0=0$. Equation~\eqref{eq:DRexact} gives a closed\hyph{}form expression for $D_R(\alpha)$ at every $0\le\alpha<1$. Both results include the classical limit. The pure quantum walk, with no classical steps, spreads ballistically and lies outside the diffusive regime.

The displacement variance is calculated exactly for each finite interval between classical steps. The calculation uses the curvature of a transfer\hyph{}operator eigenvalue. For the periodic protocol, the displacement variance gives an explicit finite sum for the diffusion coefficient. For the random protocol, averaging the displacement variance over the geometric distribution of interval lengths gives a closed\hyph{}form diffusion coefficient. 

\subsection{Temporal arrangement at matched frequency}

The exact results quantify the effect of temporal arrangement at the same average frequency of classical steps. At $\varepsilon=1-\alpha=1-r$, the quantum\hyph{}limit behavior is
\begin{equation}
D_R\sim\frac{2-\sqrt2}{2\varepsilon},
\quad D_P\sim\frac{2-\sqrt2}{4\varepsilon}.
\label{eq:diverge}
\end{equation}
Both coefficients diverge inversely with $\varepsilon$, and the ratio $D_R/D_P$ approaches two as $\varepsilon\to0$. The exact coefficients also quantify the difference between the two protocols at every matched nonzero frequency.

The difference between $D_R$ and $D_P$ near the quantum limit follows from the interval\hyph{}length statistics. For the random protocol, the interval length $\ell$ follows a geometric distribution. For the periodic protocol, the interval length is fixed at $1/\varepsilon$. At the same average frequency of classical steps, the first and second moments are
\begin{equation}
\begin{aligned}
\mathbb E_R[\ell]&=\mathbb E_P[\ell]=\frac1\varepsilon, \cr
\mathbb E_R[\ell^2]&=\frac{2-\varepsilon}{\varepsilon^2},
\quad \mathbb E_P[\ell^2]=\frac1{\varepsilon^2}.
\end{aligned}
\label{eq:interval_moments}
\end{equation}
The symbols $\mathbb E_R$ and $\mathbb E_P$ denote averages over the interval\hyph{}length distributions for the random and periodic protocols, respectively. The mean interval lengths are equal, but the second moments differ.

For a fixed interval length $\ell$, both protocols have the same displacement variance $V_\ell$. Each diffusion coefficient equals the mean displacement variance per interval divided by twice the mean interval length:
\begin{equation}
D_R=\frac{\mathbb E_R[V_\ell]}{2\mathbb E_R[\ell]},
\quad D_P=\frac{\mathbb E_P[V_\ell]}{2\mathbb E_P[\ell]}.
\end{equation}
At matched average frequencies, the denominators are equal. The difference between $D_R$ and $D_P$ therefore comes from the averages of $V_\ell$.

For long intervals, ballistic spreading of the Hadamard walk~\cite{ambainis2001one,grimmett2004weak} gives $V_\ell\sim b\ell^2$, where $b=1-1/\sqrt2$. The diffusion coefficients therefore depend on the second moments of the interval lengths as $\varepsilon\to0$:
\begin{equation}
\begin{aligned}
D_R&\sim\frac{b,\mathbb E_R[\ell^2]}{2\mathbb E_R[\ell]}
\sim\frac{b}{\varepsilon},[2pt]
D_P&\sim\frac{b,\mathbb E_P[\ell^2]}{2\mathbb E_P[\ell]}
=\frac{b}{2\varepsilon}.
\end{aligned}
\end{equation}
The ratio $\mathbb E_R[\ell^2]/\mathbb E_P[\ell^2]$ approaches two. The coefficients multiplying $1/\varepsilon$ are therefore $C_R=b$ for the random protocol and $C_P=b/2$ for the periodic protocol.

\subsection{Benchmarks for finite-mode approximations}

Comparison with the exact diffusion coefficient $D_R(\alpha)$ quantifies the error of the finite\hyph{}mode approximations. Retaining the Fourier modes $|m|\le L$ gives $D_R^{(L)}(\alpha)$ from the eigenvalue curvature of the finite transfer matrix $\mathcal W_R^{(L)}$. For $L=2$, the transfer matrix has dimension 20. The relative error is below $0.1\%$ at the values in Table~\ref{tab:approx} and about $0.072\%$ in the quantum\hyph{}limit amplitude.

The finite\hyph{}mode calculation provides a flexible approach to approximating diffusion coefficients. Changes to the coin operation or the classical step can be incorporated by modifying the transfer matrix. The calculation can therefore be extended to a broader range of quantum--classical hybrid models without requiring an exact expression for the displacement variance.

\subsection{Scope and extensions}

The exact calculation relies on the specified classical step. Each classical step removes quantum coherence and prepares the coin in an equal mixture of the two directions. The calculation also requires independence between successive interval lengths and between successive interval displacements. For a given interval length, the mean displacement must be zero, and the displacement distribution must be the same at every starting node.

Spatial structure affects both transport and collective behavior. Classical transport depends on spatial and spectral dimension~\cite{polya1921walk,hwang2010spectral,hwang2012effective,hwang2012first,hwang2013origin,kim2025spatial}. Collective spin behavior depends on dimension and topology~\cite{onsager1944crystal,dorogovtsev2008critical,jang2015ashkin,kim2021link,kim2025ashkin,kim2026heterogeneous}. Quantum transport depends on geometry and the coin operation~\cite{mackay2002higher,higuchi2014spectral,komatsu2018grover,asahara2026two}. Higher\hyph{}dimensional lattices and complex networks, including biological networks, offer settings for studying the role of spatial structure in quantum--classical hybrid transport~\cite{masuda2017random,sousa2022quantifying,miron2021transition,park2026advancing}.

Changes to the spatial structure or the classical step may invalidate the assumptions of the interval calculation. Displacements in different intervals may become statistically dependent. For a fixed interval length, the displacement distribution may also depend on the starting node. The applicability of the interval calculation must therefore be checked for each extended model.

Under the stated assumptions, the interval calculation compares the periodic and random protocols at the same average frequency of classical steps. If the displacement variance grows as $V_\ell\sim b\ell^2$ for long intervals, the ratio $D_R/D_P$ approaches two as $\varepsilon\to0$. The exact results for the one\hyph{}dimensional model provide a reference for studying the effects of spatial structure on diffusion in more general quantum--classical hybrid walks.

\begin{acknowledgments}
C.H.K. is funded by the National Research Foundation of Korea (Grant Nos. RS-2023-00279802 and RS-2026-25556154), and the KENTECH Research Grant No. KRG-2021-01-007. J.P. is funded by Helmsley Charitable Trust, the Leona M and Harry B Helmsley Charitable Trust grant (Grant No. G118500). N.H. is funded by the Brain Pool Plus Fellowship Program, which is supported by the Ministry of Science and ICT (Grant No. RS-2025-25427881) and the Korean ARPA-H Project through the Korea Health Industry Development Institute funded by the Ministry of Health and Welfare, Republic of Korea (Grant No. RS-2025-25456722).

\end{acknowledgments}

\bibliographystyle{apsrev4-2}
\bibliography{ref}

@article{polya1921walk,
  author  = {P{\'o}lya, George},
  title   = {{\"Uber eine Aufgabe der Wahrscheinlichkeitsrechnung betreffend die Irrfahrt im Stra{\ss}ennetz}},
  journal = {Math. Ann.},
  volume  = {84},
  number  = {1--2},
  pages   = {149--160},
  year    = {1921},
  doi     = {10.1007/BF01458701}
}

@article{aharonov1993quantum,
  author  = {Aharonov, Yakir and Davidovich, Luiz and Zagury, Nicim},
  title   = {Quantum random walks},
  journal = {Phys. Rev. A},
  volume  = {48},
  number  = {2},
  pages   = {1687--1690},
  year    = {1993},
  doi     = {10.1103/PhysRevA.48.1687}
}

@inproceedings{ambainis2001one,
  author    = {Ambainis, Andris and Bach, Eric and Nayak, Ashwin and Vishwanath, Ashvin and Watrous, John},
  title     = {One-dimensional quantum walks},
  booktitle = {Proceedings of the Thirty-Third Annual {ACM} Symposium on Theory of Computing},
  pages     = {37--49},
  year      = {2001},
  publisher = {Association for Computing Machinery},
  address   = {New York, NY, USA},
  doi       = {10.1145/380752.380757}
}

@inproceedings{childs2003exponential,
  author    = {Childs, Andrew M. and Cleve, Richard and Deotto, Enrico and Farhi, Edward and Gutmann, Sam and Spielman, Daniel A.},
  title     = {Exponential algorithmic speedup by a quantum walk},
  booktitle = {Proceedings of the Thirty-Fifth Annual {ACM} Symposium on Theory of Computing},
  pages     = {59--68},
  year      = {2003},
  publisher = {Association for Computing Machinery},
  address   = {New York, NY, USA},
  doi       = {10.1145/780542.780552}
}

@article{kempe2003quantum,
  author  = {Kempe, Julia},
  title   = {Quantum random walks: An introductory overview},
  journal = {Contemp. Phys.},
  volume  = {44},
  number  = {4},
  pages   = {307--327},
  year    = {2003},
  doi     = {10.1080/00107151031000110776}
}

@article{shenvi2003quantum,
  author  = {Shenvi, Neil and Kempe, Julia and Whaley, K. Birgitta},
  title   = {Quantum random-walk search algorithm},
  journal = {Phys. Rev. A},
  volume  = {67},
  number  = {5},
  pages   = {052307},
  year    = {2003},
  doi     = {10.1103/PhysRevA.67.052307}
}

@article{grimmett2004weak,
  author  = {Grimmett, Geoffrey and Janson, Svante and Scudo, Petra F.},
  title   = {Weak limits for quantum random walks},
  journal = {Phys. Rev. E},
  volume  = {69},
  number  = {2},
  pages   = {026119},
  year    = {2004},
  doi     = {10.1103/PhysRevE.69.026119}
}

@article{mohseni2008environment,
  author  = {Mohseni, Masoud and Rebentrost, Patrick and Lloyd, Seth and Aspuru-Guzik, Al{\'a}n},
  title   = {Environment-assisted quantum walks in photosynthetic energy transfer},
  journal = {J. Chem. Phys.},
  volume  = {129},
  number  = {17},
  pages   = {174106},
  year    = {2008},
  doi     = {10.1063/1.3002335}
}

@article{plenio2008dephasing,
  author  = {Plenio, M. B. and Huelga, S. F.},
  title   = {Dephasing-assisted transport: quantum networks and biomolecules},
  journal = {New J. Phys.},
  volume  = {10},
  number  = {11},
  pages   = {113019},
  year    = {2008},
  doi     = {10.1088/1367-2630/10/11/113019}
}

@article{brun2003decoherent,
  author  = {Brun, Todd A. and Carteret, Hilary A. and Ambainis, Andris},
  title   = {Quantum random walks with decoherent coins},
  journal = {Phys. Rev. A},
  volume  = {67},
  number  = {3},
  pages   = {032304},
  year    = {2003},
  doi     = {10.1103/PhysRevA.67.032304}
}

@article{brun2003transition,
  author  = {Brun, Todd A. and Carteret, Hilary A. and Ambainis, Andris},
  title   = {Quantum to classical transition for random walks},
  journal = {Phys. Rev. Lett.},
  volume  = {91},
  number  = {13},
  pages   = {130602},
  year    = {2003},
  doi     = {10.1103/PhysRevLett.91.130602}
}

@article{kendon2003decoherence,
  author  = {Kendon, Viv and Tregenna, Ben},
  title   = {Decoherence can be useful in quantum walks},
  journal = {Phys. Rev. A},
  volume  = {67},
  number  = {4},
  pages   = {042315},
  year    = {2003},
  doi     = {10.1103/PhysRevA.67.042315}
}

@article{romanelli2005decoherence,
  author  = {Romanelli, Alejandro and Siri, Rodrigo and Abal, Gerardo and Auyuanet, Adri{\'a}n and Donangelo, Ra{\'u}l},
  title   = {Decoherence in the quantum walk on the line},
  journal = {Physica A},
  volume  = {347},
  pages   = {137--152},
  year    = {2005},
  doi     = {10.1016/j.physa.2004.08.070}
}

@article{kendon2007review,
  author  = {Kendon, Viv},
  title   = {Decoherence in quantum walks---A review},
  journal = {Math. Struct. Comput. Sci.},
  volume  = {17},
  number  = {6},
  pages   = {1169--1220},
  year    = {2007},
  doi     = {10.1017/S0960129507006354}
}

@article{romanelli2007measurements,
  author  = {Romanelli, Alejandro},
  title   = {Measurements in the {L\'e}vy quantum walk},
  journal = {Phys. Rev. A},
  volume  = {76},
  number  = {5},
  pages   = {054306},
  year    = {2007},
  doi     = {10.1103/PhysRevA.76.054306}
}

@article{abal2008generic,
  author  = {Abal, Gerardo and Donangelo, Ra{\'u}l and Severo, Fabricio and Siri, Rodrigo},
  title   = {Decoherent quantum walks driven by a generic coin operation},
  journal = {Physica A},
  volume  = {387},
  number  = {1},
  pages   = {335--345},
  year    = {2008},
  doi     = {10.1016/j.physa.2007.08.058}
}

@article{zhang2008limiting,
 author = {Zhang, Kai},
 title = {Limiting distribution of decoherent quantum random walks},
 journal = {Physical Review A},
 volume = {77}, pages = {062302}, year = {2008},
 doi = {10.1103/PhysRevA.77.062302}
}

@article{annabestani2010decoherence,
  author  = {Annabestani, Mostafa and Akhtarshenas, Seyed Javad and Abolhassani, Mohamad Reza},
  title   = {Decoherence in a one-dimensional quantum walk},
  journal = {Phys. Rev. A},
  volume  = {81},
  number  = {3},
  pages   = {032321},
  year    = {2010},
  doi     = {10.1103/PhysRevA.81.032321}
}

@article{whitfield2010quantum,
  author  = {Whitfield, James D. and Rodr{\'i}guez-Rosario, C{\'e}sar A. and Aspuru-Guzik, Al{\'a}n},
  title   = {Quantum stochastic walks: A generalization of classical random walks and quantum walks},
  journal = {Phys. Rev. A},
  volume  = {81},
  number  = {2},
  pages   = {022323},
  year    = {2010},
  doi     = {10.1103/PhysRevA.81.022323}
}

@article{ahlbrecht2011asymptotic,
  author  = {Ahlbrecht, Andre and Vogts, Holger and Werner, Albert H. and Werner, Reinhard F.},
  title   = {Asymptotic evolution of quantum walks with random coin},
  journal = {J. Math. Phys.},
  volume  = {52},
  number  = {4},
  pages   = {042201},
  year    = {2011},
  doi     = {10.1063/1.3575568}
}

@article{schreiber2011decoherence,
  author  = {Schreiber, Andrea and Cassemiro, Katiuscia N. and Poto{\v{c}}ek, V{\'a}clav and G{\'a}bris, Aur{\'e}l and Jex, Igor and Silberhorn, Christine},
  title   = {Decoherence and disorder in quantum walks: From ballistic spread to localization},
  journal = {Phys. Rev. Lett.},
  volume  = {106},
  number  = {18},
  pages   = {180403},
  year    = {2011},
  doi     = {10.1103/PhysRevLett.106.180403}
}

@article{ahlbrecht2012spatiotemporal,
  author  = {Ahlbrecht, Andre and Cedzich, Christopher and Matjeschk, Robert and Scholz, Volkher B. and Werner, Albert H. and Werner, Reinhard F.},
  title   = {Asymptotic behavior of quantum walks with spatio-temporal coin fluctuations},
  journal = {Quantum Inf. Process.},
  volume  = {11},
  number  = {5},
  pages   = {1219--1249},
  year    = {2012},
  doi     = {10.1007/s11128-012-0389-4}
}

@article{mackay2002higher,
  author  = {Mackay, Troy D. and Bartlett, Stephen D. and Stephenson, Leigh T. and Sanders, Barry C.},
  title   = {Quantum walks in higher dimensions},
  journal = {J. Phys. A},
  volume  = {35},
  number  = {12},
  pages   = {2745--2753},
  year    = {2002},
  doi     = {10.1088/0305-4470/35/12/304}
}

@article{higuchi2014spectral,
  author  = {Higuchi, Yusuke and Konno, Norio and Sato, Iwao and Segawa, Etsuo},
  title   = {Spectral and asymptotic properties of {Grover} walks on crystal lattices},
  journal = {J. Funct. Anal.},
  volume  = {267},
  number  = {11},
  pages   = {4197--4235},
  year    = {2014},
  doi     = {10.1016/j.jfa.2014.09.003}
}

@article{komatsu2018grover,
  author  = {Komatsu, Takashi and Tate, Tatsuya},
  title   = {Eigenvalues of quantum walks of {Grover} and {Fourier} types},
  journal = {J. Fourier Anal. Appl.},
  volume  = {25},
  number  = {4},
  pages   = {1293--1318},
  year    = {2019},
  doi     = {10.1007/s00041-018-9630-6}
}

@article{asahara2026two,
  author  = {Asahara, Keisuke and Funakawa, Daiju and Seki, Motoki and Suzuki, Akito},
  title   = {Two-dimensional quantum central limit theorem by quantum walks},
  journal = {Phys. Rev. A},
  volume  = {113},
  number  = {2},
  pages   = {022439},
  year    = {2026},
  doi     = {10.1103/fs8d-h2z8}
}

@article{hwang2010spectral,
  author  = {Hwang, S. and Yun, C.-K. and Lee, D.-S. and Kahng, B. and Kim, D.},
  title   = {Spectral dimensions of hierarchical scale-free networks with weighted shortcuts},
  journal = {Phys. Rev. E},
  volume  = {82},
  number  = {5},
  pages   = {056110},
  year    = {2010},
  doi     = {10.1103/PhysRevE.82.056110}
}

@article{hwang2012effective,
  author  = {Hwang, S. and Lee, D.-S. and Kahng, B.},
  title   = {Effective trapping of random walkers in complex networks},
  journal = {Phys. Rev. E},
  volume  = {85},
  number  = {4},
  pages   = {046110},
  year    = {2012},
  doi     = {10.1103/PhysRevE.85.046110}
}

@article{hwang2012first,
  author  = {Hwang, S. and Lee, D.-S. and Kahng, B.},
  title   = {First passage time for random walks in heterogeneous networks},
  journal = {Phys. Rev. Lett.},
  volume  = {109},
  number  = {8},
  pages   = {088701},
  year    = {2012},
  doi     = {10.1103/PhysRevLett.109.088701}
}

@article{hwang2013origin,
  author  = {Hwang, S. and Lee, D.-S. and Kahng, B.},
  title   = {Origin of the hub spectral dimension in scale-free networks},
  journal = {Phys. Rev. E},
  volume  = {87},
  number  = {2},
  pages   = {022816},
  year    = {2013},
  doi     = {10.1103/PhysRevE.87.022816}
}

@article{masuda2017random,
  author  = {Masuda, Naoki and Porter, Mason A. and Lambiotte, Renaud},
  title   = {Random walks and diffusion on networks},
  journal = {Phys. Rep.},
  volume  = {716--717},
  pages   = {1--58},
  year    = {2017},
  doi     = {10.1016/j.physrep.2017.07.007}
}

@article{kim2025spatial,
  author  = {Kim, Cook Hyun and Kahng, B.},
  title   = {From spatial to spectral: Network renormalization via dynamical correlations},
  journal = {Chaos Solitons Fractals},
  volume  = {201},
  pages   = {117398},
  year    = {2025},
  doi     = {10.1016/j.chaos.2025.117398}
}

@article{miron2021transition,
  author  = {Miron, Philippe and Beron-Vera, Francisco J. and Helfmann, Luzie and Koltai, P{\'e}ter},
  title   = {Transition paths of marine debris and the stability of the garbage patches},
  journal = {Chaos},
  volume  = {31},
  number  = {3},
  pages   = {033101},
  year    = {2021},
  doi     = {10.1063/5.0030535}
}

@article{sousa2022quantifying,
  author  = {Sousa, Sandro and Nicosia, Vincenzo},
  title   = {Quantifying ethnic segregation in cities through random walks},
  journal = {Nat. Commun.},
  volume  = {13},
  pages   = {5809},
  year    = {2022},
  doi     = {10.1038/s41467-022-33344-3}
}

@article{park2026advancing,
  author  = {Park, Jaesub and Hwang, Woochang and Lee, Seokjun and Lee, Hyun Chang and MacMahon, M{\'e}abh and Zilbauer, Matthias and Han, Namshik},
  title   = {Advancing understanding of long {COVID} pathophysiology through quantum walk-based network analysis},
  journal = {Bioinform. Adv.},
  volume  = {6},
  number  = {1},
  pages   = {vbag050},
  year    = {2026},
  doi     = {10.1093/bioadv/vbag050}
}

@article{onsager1944crystal,
  author  = {Onsager, Lars},
  title   = {Crystal statistics. {I}. A two-dimensional model with an order-disorder transition},
  journal = {Phys. Rev.},
  volume  = {65},
  number  = {3--4},
  pages   = {117--149},
  year    = {1944},
  doi     = {10.1103/PhysRev.65.117}
}

@article{dorogovtsev2008critical,
  author  = {Dorogovtsev, S. N. and Goltsev, A. V. and Mendes, J. F. F.},
  title   = {Critical phenomena in complex networks},
  journal = {Rev. Mod. Phys.},
  volume  = {80},
  number  = {4},
  pages   = {1275--1335},
  year    = {2008},
  doi     = {10.1103/RevModPhys.80.1275}
}

@article{jang2015ashkin,
  author  = {Jang, S. and Lee, J. S. and Hwang, S. and Kahng, B.},
  title   = {{Ashkin}-{Teller} model and diverse opinion phase transitions on multiplex networks},
  journal = {Phys. Rev. E},
  volume  = {92},
  number  = {2},
  pages   = {022110},
  year    = {2015},
  doi     = {10.1103/PhysRevE.92.022110}
}

@article{kim2021link,
  author  = {Kim, Cook Hyun and Jo, Minjae and Lee, J. S. and Bianconi, Ginestra and Kahng, B.},
  title   = {Link overlap influences opinion dynamics on multiplex networks of {Ashkin}-{Teller} spins},
  journal = {Phys. Rev. E},
  volume  = {104},
  number  = {6},
  pages   = {064304},
  year    = {2021},
  doi     = {10.1103/PhysRevE.104.064304}
}

@article{kim2025ashkin,
  author  = {Kim, Cook Hyun and Choi, Hoyun and Jung, Joonsung and Kahng, B.},
  title   = {{Ashkin}-{Teller} model with antiferromagnetic four-spin interactions: Interference effect between two conflicting issues},
  journal = {Chaos Solitons Fractals},
  volume  = {199},
  pages   = {116787},
  year    = {2025},
  doi     = {10.1016/j.chaos.2025.116787}
}

@article{kim2026heterogeneous,
  author  = {Kim, Cook Hyun and Kahng, B.},
  title   = {Heterogeneous network topology induces the {Widom} line},
  journal = {Phys. Rev. E},
  volume  = {113},
  number  = {5},
  pages   = {055410},
  year    = {2026},
  doi     = {10.1103/qvrb-yg3g}
}
\end{document}